\documentclass{aa}  

\usepackage{graphicx}
\usepackage{txfonts}
\usepackage{xcolor}
\usepackage{amsmath}
\usepackage{nccmath}
\usepackage{gensymb}
\usepackage{ragged2e}
\usepackage[colorlinks=true,linkcolor=blue,citecolor={blue}]{hyperref}%
\newcommand{\Nstar}{HD~984}	
\newcommand{\NTstar}{HIP~115119}	
\newcommand{\NMstar}{HIP~95771}

\newcommand{\Ncomp}{HD~984~B}

\newcommand{\pRT}{\texttt{petitRADTRANS}}

\begin{document}

   \title{Fresh view of the hot brown dwarf \Ncomp\ through high-resolution spectroscopy}

   \author{J.~C.~Costes\inst{1}\thanks{E-mail: \href{jean.costes@lam.fr}{jean.costes@lam.fr}}
          \and
          J.~W.~Xuan\inst{2}
          \and
          A.~Vigan\inst{1}
          \and
          J.~Wang\inst{2,3}
          \and
          V.~D'Orazi\inst{4,5}
          \and
          P.~Mollière\inst{6}
          \and
          A.~Baker\inst{2}
          \and
          R.~Bartos\inst{7}
          \and
          G.~A.~Blake\inst{8}
          \and
          B.~Calvin\inst{2,9}
          \and
          S.~Cetre\inst{10}
          \and
          J.~Delorme\inst{10}
          \and
          G.~Doppmann\inst{10}
          \and
          D.~Echeveri\inst{2}
          \and
          L.~Finnerty\inst{9}
          \and
          M.~P.~Fitzgerald\inst{9}
          \and
          C.~Hsu\inst{3}
          \and
          N.~Jovanovic\inst{2}
          \and
          R.~Lopez\inst{9}
          \and
          D.~Mawet\inst{2,7}
          \and
          E.~Morris\inst{11}
          \and
          J.~Pezzato\inst{2}
          \and
          C.~L.~Phillips\inst{13}
          \and
          J.~Ruffio\inst{12}
          \and
          B.~Sappey\inst{12}
          \and
          A.~Schneeberger\inst{1}
          \and
          T.~Schofield\inst{2}
          \and
          A.~J.~Skemer\inst{11}
          \and
          J.~K.~Wallace\inst{7}
          \and
          J.~Wang\inst{13}
          }

   \institute{Aix Marseille Univ, CNRS, CNES, LAM, Marseille, France
             \and
             Department of Astronomy, California Institute of Technology, Pasadena, CA 91125, USA
             \and
             Center for Interdisciplinary Exploration and Research in Astrophysics (CIERA) and Department of Physics and Astronomy, Northwestern University, Evanston, IL 60208, USA
             \and
             Department of Physics, University of Rome Tor Vergata, via della Ricerca Scientifica 1, 00133, Rome, Italy
             \and
             INAF Osservatorio Astornomico di Padova, vicolo dell'Osservatorio 5, 35122, Padova, Italy
             \and
             Max Planck Institut für Astronomie, Königstuhl 17, D-69117 Heidelberg, Germany
             \and
             Jet Propulsion Laboratory, California Institute of Technology, 4800 Oak Grove Dr., Pasadena, CA 91109, USA
             \and
             Division of Geological \& Planetary Sciences, California Institute of Technology, Pasadena, CA 91125, USA
             \and
             Department of Physics \& Astronomy, 430 Portola Plaza, University of California, Los Angeles, CA 90095, USA
             \and
             W. M. Keck Observatory, 65-1120 Mamalahoa Hwy, Kamuela, HI, USA
             \and
             Department of Astronomy \& Astrophysics, University of California, Santa Cruz, CA95064, USA
             \and
             Center for Astrophysics and Space Sciences, University of California, San Diego, La Jolla, CA 92093
             \and
             Department of Astronomy, The Ohio State University, 100 W 18th Ave, Columbus, OH 43210 USA
             }

   \date{}

 
  \abstract
   {High-resolution spectroscopy has the potential to drive a better understanding of the atmospheric composition, physics, and dynamics of young exoplanets and brown dwarfs, bringing clear insights into the formation channel of individual objects.}
   {Using the Keck Planet Imager and Characterizer (KPIC; $\mathrm{R \approx 35,000}$), we aim to characterize a young brown dwarf \Ncomp. By measuring its C/O and $\mathrm{^{12}CO/^{13}CO}$ ratios, we expect to gain new knowledge about its origin by confirming the difference in the formation pathways between brown dwarfs and super-Jupiters.}
   {We analysed the KPIC high-resolution spectrum (2.29–2.49 $\mu$m) of \Ncomp\ using an atmospheric retrieval framework based on nested sampling and \pRT, using both clear and cloudy models.}
   {Using our best-fit model, we find $\rm{C/O}=0.50 \pm 0.01$ (0.01 is the statistical error) for \Ncomp\, which agrees with that of its host star within 1$\sigma$ ($0.40 \pm 0.20$). We also retrieve an isotopolog $\mathrm{^{12}CO/^{13}CO}$ ratio of $98^{+20}_{-25}$ in its atmosphere, which is similar to that of the Sun. In addition, \Ncomp\ has a substellar metallicity with $[$Fe/H$] = -0.62^{+0.02}_{-0.02}$. Finally, we find that most of the retrieved parameters are independent of our choice of retrieval model.}
   {From our measured C/O and $\mathrm{^{12}CO/^{13}CO}$, the favored formation mechanism of \Ncomp\ seems to be via gravitational collapse or disk instability and not core accretion, which is a favored formation mechanism for giant exoplanets with $m<13~\mathrm{M_{Jup}}$ and semimajor axis between 10 and 100~au. However, with only a few brown dwarfs with a measured $\mathrm{^{12}CO/^{13}CO}$ ratio, similar analyses using high-resolution spectroscopy will become essential in order to determine planet formation processes more precisely.}


   \maketitle


\section{Introduction}
Over the past 20~years, several dozen substellar companions have been directly imaged in the near-infrared using large ground-based telescopes. These companions are generally massive (from 2 to $70~\mathrm{M_{Jup}}$), with large orbital separations from their parent star (from 3 to 1000~au; for a review, see e.g., \citealt{Bowler2016, Currie2023}). This population of massive distant objects is of high interest because they challenge several scenarios of giant planet and star formation. Recent large direct-imaging surveys have placed tight constraints on the population of these objects \citep{Nielsen2019, Vigan2021} and showed that multiple formation channels likely cause the population of imaged companions. Between 10 and 100~au, giant planets (defined as planets with $m<13~\mathrm{M_{Jup}}$) have a higher occurrence rate than more massive brown dwarfs ($m>13~\mathrm{M_{Jup}}$). In addition, giant exoplanets preferentially occur around higher-mass stars and likely formed via core accretion \citep{Pollack1996, Alibert2005}. Brown dwarf companions, however, represent the low-mass tail of the stellar binary population and likely formed via disk instability \citep{Boss1997, Kratter201}. Differences in formation mechanisms were also reinforced by \cite{Bowler2020}, \cite{DoO2023}, and \cite{Nagpal2023}, who showed that giant planets at wide separations (between $5-100$~au) had a low-eccentricity distribution similar to that of close-in giant planets ($<1$~au). This is in contrast to the brown dwarf eccentricity distribution, which is more similar to the stellar binary population. While recent surveys gained information about the population as a whole, one of the main challenges is to determine the formation channel of individual objects.

Measuring the carbon-to-oxygen ratio (C/O) of a planet has been suggested to be a powerful tracer of the location in which a planet may have formed in the protoplanetary disk (e.g., \citealt{Oberg2011, Cridland2020, Chachan2023}). This analysis essentially relies on comparisons between the planetary C/O to the C/O predicted for the solid and vapor phases of the disk. An analysis like this was performed for the HR~8799 system \citep{Konopacky2013, Molliere2020}, for instance, and led to the conclusion that the planets likely formed outside the $\mathrm{CO_{2}}$ and CO snowlines. A similar analysis was performed for $\beta$~Pictoris~b \citep{GRAVITY2020}, and the results suggested a formation through core-accretion with strong planetesimal enrichment. 
Another more recent suite of tracers that was considered are isotopolog ratios, with a possible difference in the $\mathrm{^{12}CO/^{13}CO}$ ratio between giant planets and brown dwarfs (e.g., \citealt{Zhang2021a}). A final distinct approach is the idea that the rotational velocity of a giant exoplanet bears the signature of its initial angular momentum that is accumulated during the gas-accretion phase, and that different formation channels may show measurable differences in the $v\sin i$ \citep{Bryan2016, Bowler2023}. This approach was tested by \citet{Bryan2018}, but the results did not show statistically measurable differences between the population of companions and of isolated objects. Their sample, however, was very small, and a more thorough exploration of this approach is necessary by increasing the number of $v\sin i$ measurements for young substellar companions.

High-resolution spectroscopy of directly imaged companions is the method of choice for characterizing their orbits, spin, and compositions. This method relies on the known (or assumed) spectral signature of the companion and on techniques such as a cross-correlation analysis for isolating the signal. This approach has proven efficient at detecting the molecular signature of known companions and strongly constrains the detailed composition of planetary atmospheres \citep[e.g.,][]{Konopacky2013, Brogi2019}. The radial velocity (RV) of the companion can be measured by quantifying the Doppler shift of molecular absorption lines, while the planetary spin can be inferred from their rotational broadening \citep{Snellen2014, Bryan2018}. 

To date, most high-resolution spectroscopic observations of of directly imaged companions assisted by adaptive optics (AO) have been limited to bright companions at large angular separations from their host star. This limitation is explained by the fact that the host star overwhelms the signal of the planet. Therefore, only a few directly imaged companions within $1''$ have been observed with high-resolution spectroscopy so far (e.g., \citealt{Snellen2014, Wang2018, Zhang2021b, Xuan2022, Landman2024}). KPIC is a new suite of instrument upgrades at Keck II that combines high-contrast imaging techniques with high-resolution spectroscopy \citep{Mawet2017, Delorme2021}. KPIC is composed of a single-mode fibre injection unit that feeds light into the upgraded Near InfraRed Spectrograph (NIRSPEC; \citealt{Martin2018, Lopez2020}) and enables high-resolution spectroscopy at $\mathrm{R\approx35,000}$. With KPIC, it is therefore possible to further distinguish between the star and the planet (e.g., \citealt{Delorme2021, Wang2023}), which allows us to observe and characterize fainter and closer-in directly imaged exoplanets. For instance, \cite{Wang2021a} and \cite{Xuan2022} described that KPIC is able to detect molecular lines and to measure the rotational line broadening of companions at high contrast ($\Delta\mathrm{K}\approx11$ and $8$, respectively) and small separation ($\approx0.4''$ and $0.6''$) for the HR~8799 system and HD~4747~B, allowing for a better characterization of the companions. 

We use KPIC data here to study the characteristics and composition of the atmosphere of the young brown dwarf companion \Ncomp. In Sect.~\ref{Observations}, we present the system  and the observations made with KPIC. In Sect.~\ref{Data_reduction}, we then describe the data reduction procedure as well as the preliminary results. In Sect.~\ref{Spectral_Analysis}, we detail the full analysis of our spectroscopic data using \pRT. Finally, we discuss these results and present future prospects in Sect.~\ref{conclusion}.

\section{Observations}\label{Observations}
\subsection{System }\label{Host_Star}

\Nstar\ is a young ($30 - 200$~Myr) F7 star hosting a low-mass companion that was first detected via high-contrast imaging \citep{Meshkat2015}. \Nstar\ has a mass of $\sim$1.2~$\mathrm{M_{\odot}}$, a temperature of $6315\pm89$~K \citep{Casagrande2011, Meshkat2015}, and is located at a distance of $45.6$~parsec \citep{Gaia2021}. \Nstar\ is one of fastest rotating stars, with a stellar rotational period of $\mathrm{P_{rot}}<1.6$~d that is measured from its $v\sin i = 39.3\pm1.5~\mathrm{km~s}^{-1}$ and radius $\mathrm{R_s}=1.247\pm0.053~\mathrm{R_{\odot}}$ \citep{Meshkat2015}. Only   $1\%$ of all stars rotate this fast. \Nstar\ was first reported to be part of the 30~Myr old Columba association \citep{Zuckerman2011}. However, recent studies showed that a convincing kinematic match to Columba or other young comoving stars is lacking \citep{Meshkat2015, Franson2022}. 

Because the stellar rotation is so high, and in order to acquire crucial information regarding stellar parameters and abundances for key species such as iron, carbon, and oxygen, we used three spectra from the Fiber-fed Extended Range Optical Spectrograph (FEROS) (nominal resolution $\mathrm{R=48000}$; \citealt{kaufer1999}) that are available in the ESO archive\footnote{programs: 072.A-9006(A), 077.C-0573(A)}. The spectra provide wavelength coverage from $350 - 920$~nm with a median signal-to-noise ratio (S/N) per pixel of 250. After shifting each spectrum to the rest frame through a cross-correlation with synthetic templates, we computed a median spectrum using the Python packages in {\tt iSpec} \citep{blanco-cuaresma2019}. We estimated $\mathrm{T_{eff}}$ by fitting the stellar H$_\alpha$ profile using the code Spectroscopy Made Easy (SME; \citealt{piskunov2017}), assuming solar metallicity and $\mathrm{log}~g=4.3$~dex and found $\mathrm{T_{eff}=6380\pm75}$~K. Furthermore, we exploited the photometric magnitudes and colors from \textit{Gaia} DR3 and 2MASS (see Table~\ref{tab:HD984}), and used the calibration by \cite{Casagrande2021} to infer the photometric effective temperature, which we derive to be $\mathrm{T_{eff}=6272\pm82}$~K (with the error as given with Monte Carlo methods; see \citealt{Casagrande2021} for further details). We then took the average value of the two estimates as our best-fit temperature determination, where $\mathrm{T_{eff}=6326}$~K is a conservative uncertainty of $\pm~80$~K. 

Adopting this $\mathrm{T_{eff}}$ value, the bolometric correction for G magnitudes by \cite{Casagrande2018}, a mass of $\mathrm{M=1.2~M_\odot}$ \citep{Meshkat2015}, and the \textit{Gaia} parallax, we calculated the surface gravity as
\begin{equation}
   \mathrm{log}~g~=~4.44~+~\mathrm{log~\frac{M}{M_\odot}}~-~2.5~\mathrm{log~\frac{L}{L_\odot}}~+~4~\mathrm{log~\frac{T_{eff}}{T_{eff\odot}}}~,
\end{equation} which results in $4.38~\pm~0.06$~dex. The error takes the uncertainties on mass, magnitudes, and $\mathrm{T_{eff}}$ into account. For the microturbulence velocity, we exploited the relation $\mathrm{V_t=2.13 - 0.23 \times \log g}$ \citep{Kirby2009} and retrieved a value of $1.2~\pm~0.10$~km~s$^{-1}$, which we adopted throughout our analysis. 

\begin{figure}
    \centering
    \includegraphics[width=0.5\textwidth]{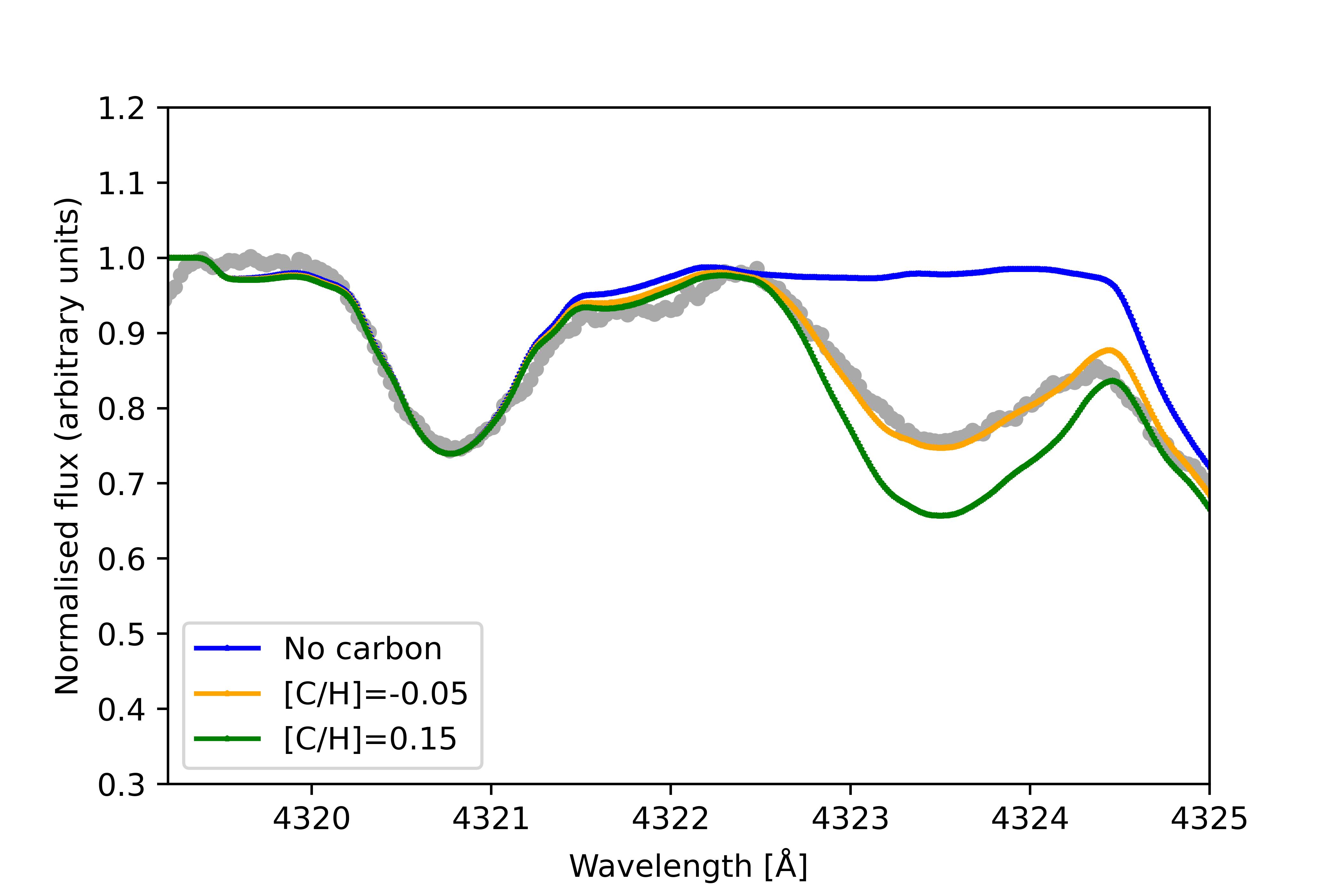}
    \caption{Spectral synthesis in the CH region at 4300 \AA\ for \Nstar.}
    \label{fig:carbon_star}
\end{figure}

Finally, the iron abundance was derived with the Python wrapper {\tt pymoogi} by M. Adamow\footnote{\url{https://github.com/madamow/pymoogi}} of the Local Thermodynamic Equilibrium (LTE) radiative transfer code {\sc moog} by C. Sneden (\citeyear{Sneden1973}, 2019 version). We selected strong, relatively isolated Fe {\sc i} lines from the \textit{Gaia}-ESO linelist \citep{Heiter2021} and computed the spectral synthesis using the driver $synth$. We obtained a metallicity of $\mathrm{[Fe/H]}=-0.01\pm0.12$~dex. The errors include the line-by-line scatter and the uncertainties on the stellar parameters. We refer to \citealt{dorazi2020} for further details of the error analysis and calculation. The carbon abundance was determined by analyzing the CH G-band ($4290-4320$~\AA), using a line list provided by B. Plez through private communication (see Figure~\ref{fig:carbon_star}). Taking advantage of the large spectral coverage that FEROS spectra offer, we obtained the oxygen content from the permitted O{ \sc i} triplet at 7770~\AA, for which we applied the Non-Local Thermodynamic Equilibrium (NLTE) corrections by \cite{Amarsi2019}. The final abundances are $\mathrm{[C/H]}=-0.05\pm0.10$~dex and $\mathrm{[O/H]}=0.09~\pm~0.20$~dex (the average abundance in LTE would be $\mathrm{[O/H]_{LTE}}=0.40$~dex, which is not compatible with a thin-disk abundance pattern). Our analysis suggests that within the observational uncertainties, all abundances conform to solar values and have a C/O ratio of $0.40\pm0.20$. We note that our uncertainties in the C/O ratio are rather large, but this is expected for rapidly rotating stars like this. The uncertainties primarily arise because it is difficult to accurately determine the best fit for spectral lines, which are broadened because of the high stellar rotational velocities. Additionally, substantial uncertainties linked to NLTE corrections for oxygen. The permitted triplet of oxygen at 777~nm in particular is affected by these corrections. Table~\ref{tab:HD984} summarizes the properties of the host star.

\begin{table}
    \centering
    \caption{ for \Nstar}
    \begin{tabular}{lcc} 
    \hline\hline
    \rule{0pt}{2.5ex}Property   &       Value           &References\\[.5ex]
    \hline
    \multicolumn{3}{l}{\rule{0pt}{2.5ex}Astrometric }\\
    R.A.                & $00^\mathrm{h}14^\mathrm{m}10^\mathrm{s}.3$                   & 1       \\
    Dec                 & $-07^{\circ}11'56''.8$                        & 1       \\
    \textit{Gaia} source I.D. & 2431157720981843200    & 1 \\
    $\mu_{{\rm R.A.}}$ ($\mathrm{mas~y}^{-1}$) & $104.775\pm0.036$ & 1 \\
    $\mu_{{\rm Dec.}}$ ($\mathrm{mas~y}^{-1}$) & $-68.016\pm0.022$ & 1 \\
    parallax (mas)      & $21.877\pm0.025$              & 1     \\[2ex]
    \multicolumn{3}{l}{Photometric }\\
    V (mag)             & $7.32\pm0.01$         & 2 \\
    B (mag)             & $7.82\pm0.02$         & 2 \\
    G (mag)             & $7.208\pm0.003$       & 1 \\
    J (mag)             & $6.402\pm0.023$               & 3     \\
    H (mag)             & $6.170\pm0.038$               & 3     \\
    K (mag)             & $6.073\pm0.021$               & 3     \\
    BP (mag)            & $7.460$               & 1     \\
    RP (mag)            & $6.791$               & 1     \\[2ex]
    \multicolumn{3}{l}{Derived Properties}\\
    Spectral type       & F7V               & 4 \\
    T$_{\rm eff}$ (K)   & $6326\pm80$       & 9 \\
    $[$Fe/H$]$          & $-0.01\pm0.12$    & 9 \\
    $[$O/H$]$           & $0.09\pm0.20$       & 9 \\
    $[$C/H$]$           & $-0.05\pm0.10$    & 9 \\
    C/O                 & $0.40\pm0.20$     & 9 \\
    $v\sin i$ ($\mathrm{km~s}^{-1}$)         & $39.3\pm1.5$       & 6 \\
    $\mathrm{log(R'_{HK})}$     &    $-4.33$   & 7 \\
    log $g$             & $4.38\pm0.06$       & 9 \\
    V$_\mathrm{t}$ ( $\mathrm{km~s}^{-1}$) &   $1.12\pm0.10$ & 9\\
    $\mathrm{M_{s}}$ ($\mathrm{M_{\odot}}$)       & $1.20\pm0.06$    & 8 \\
    $\mathrm{R_{s}}$ ($\mathrm{R_{\odot}}$)       & $1.247\pm0.053$  & 8 \\
    Age (Myr)           & $30-200$     & 8 \\
    Distance (pc)       & $45.631\pm0.055$        & 1 \\[.5ex]
    \hline
    \end{tabular}
    \tablebib{(1)~\cite{Gaia2021}; (2)~\cite{Hog2000}; (3)~\cite{Cutri2003}; (4)~\cite{Houk1999}; (5)~\cite{Casagrande2011}; (6)~\cite{Zuniga2021}; (7)~\cite{Boro2018}; (8)~\cite{Meshkat2015}; (9)~This work}
    \label{tab:HD984}
\end{table}

The brown dwarf companion to \Nstar\ was first detected via high-contrast imaging \citep{Meshkat2015}. \Ncomp\ has a spectral type of $\mathrm{M}6.5\pm1.5$ \citep{Johnson2017} at a contrast of $\Delta\mathrm{H}=6.43$~mag. A dynamical mass of $61\pm4~\mathrm{M_{Jup}}$ was measured for \Ncomp\ from a joint orbit fit of the relative astrometry, proper motions, and RVs \citep{Franson2022}, which also yielded a period of $\mathrm{P}=140^{+50}_{-30}$~yr, a high eccentricity of $\mathrm{e}=0.76\pm0.05$, and a semimajor axis of $\mathrm{a}~=~28^{+7}_{-4}$~au. Table~\ref{tab:HD984B} summarizes the properties of the companion.

\begin{table}
    \centering
    \caption{Properties for \Ncomp}
    \begin{tabular}{lcc} 
    \hline\hline
    \rule{0pt}{2.5ex}Property   &       Value           &References\\[.5ex]
    \hline
    \multicolumn{3}{l}{\rule{0pt}{2.5ex}Photometric Properties}\\
    J (mag)     & $13.28\pm0.06$      & 1 \\
    H (mag)     & $12.60\pm0.05$      & 1 \\
    K (mag)     & $12.20\pm0.04$    & 2 \\[2ex]
    \multicolumn{3}{l}{Derived Properties}\\
    Spectral type   & M$6.5\pm1.5$      & 1 \\
    Separation (mas) & $201.6\pm0.4$ & 3 \\
    Mass ($\mathrm{M_{Jup}}$)   & $61\pm4$  & 2 \\
    T$_{\rm eff}$ (K)       & $2730^{+120}_{-180}$      & 1 \\
    a (au)      & $28^{+7}_{-4}$      & 2 \\
    e           & $0.76\pm0.05$      & 2 \\
    i ($^{\circ}$)        & $120.8^{+1.8}_{-1.6}$      & 2 \\
    P (yr)      & $140^{+50}_{-30}$      & 2 \\[2ex]
    \multicolumn{3}{l}{Properties from this work}\\
    RV ($\mathrm{km~s}^{-1}$) & $-25.02^{+0.02}_{-0.03}$ & 4 \\
    $v\sin i$ ($\mathrm{km~s}^{-1}$) & $12.72^{+0.03}_{-0.02}$ & 4 \\
    $[$Fe/H$]$ & $-0.62 \pm 0.02$ & 4 \\
    C/O & $0.50 \pm 0.01$ & 4 \\
    $\mathrm{^{12}CO/^{13}CO}$ & $98^{+20}_{-25}$ & 4 \\[.5ex]
    \hline
    \end{tabular}
    \tablebib{(1)~\cite{Johnson2017}; (2)~\cite{Franson2022}; (3)~\cite{Meshkat2015}; (4)~This work}
    \label{tab:HD984B}
\end{table}

\subsection{KPIC observations of \Ncomp}
We observed \Nstar\ on UT 2022 August 8 using KPIC. We obtained nine spectral orders in K~band, ranging from approximately 1.94 to 2.49~$\mu$m. Similar to the observing strategy described in \cite{Wang2021a}, the host star was first observed twice using the four fibers for calibration purposes with an exposure of 60~s. From these exposures, we measured the end-to-end throughput from the top of the atmosphere to the detector for each fiber. Based on the fiber with the best throughput ($\sim 4\%$, designated as the primary science fiber), the tip/tilt mirror on the fiber injection unit (FIU) was the used to offset the star from the fiber bundle and to place the companion of interest on the primary science fiber. The offset amplitude and direction were computed using the orbital prediction tool \texttt{whereistheplanet}\footnote{\url{http://whereistheplanet.com}} \citep{Wang2021b}.

Two exposures were acquired for \Ncomp\ with an integration time of 600~s each so that the read-out noise would be negligible. By taking the difference in integration time between the host star and its companion into account, we measured a companion flux corresponding to $\mathrm{\Delta K = 7.1 \pm 0.3~mag}$, which is within 3$\sigma$ of the expected $\mathrm{\Delta K = 6.13 \pm 0.05~mag}$ reported in the literature (see Tables~\ref{tab:HD984} and \ref{tab:HD984B}). In addition to these spectra, an M giant star (\NMstar) and a telluric standard star (\NTstar) were also observed for calibration and wavelength solution purposes. Twelve exposures of 1.5~s each and five exposures of 60~s each were taken for \NMstar\ and \NTstar, respectively. A summary of the different observations taken during the night are presented in Table~\ref{tab:observations}.

\begin{table}[hbt!]
    \centering
    \caption{Summary of the observations taken on UT 2022 August 8}
    \begin{tabular}{lccc}
    \hline\hline
    \rule{0pt}{2.5ex}Target & Number of & Exposure & Purpose \\
     & observations &  time (s) &  \\[.5ex]
    \hline
    \rule{0pt}{2.5ex}\Nstar & 2 & 60 & Science \\
    \Ncomp & 2 & 600 & Science \\
    \NMstar & 12 & 1.5 & Wavelength \\
    \NTstar & 5 & 60 & Calibration~$^{(a)}$ \\[.5ex]
    \hline
    \end{tabular} 
   \tablefoot{$^{(a)}$~\NTstar\ was used in the calibration to measure the trace and to calculate the transmission.}
    \label{tab:observations}
\end{table}

\section{Data reduction}\label{Data_reduction}

\subsection{Raw data reduction}\label{Raw_Data_Reduction}
We followed the procedure described in \cite{Wang2021a} to extract spectra from the raw data. This procedure has been implemented in a public Python pipeline\footnote{\url{https://github.com/kpicteam/kpic pipeline}} by the KPIC team. First, the thermal background from the images as well as the identified bad pixels were removed using the combined instrument thermal background frames. Then, the trace of each of the four science fibers in each of the nine orders was measured using the data from the telluric standard star \NTstar. This star was mainly chosen for its spectral type (A0V) because it shows only a few stellar lines in the K~band and has a similar elevation (air mass) as our target \Nstar. A 1D Gaussian was fitted to the cross-section of the trace in each column of each order to determine the position and standard deviation of the Gaussian line-spread function (LSF). To mitigate measurement noise and biases from telluric lines, the measurements of the position and standard deviations were smoothed by fitting a cubic spline. 

For every frame, the 1D spectra were then extracted in each column of each order. To correct for the imperfect background subtraction, we additionally subtracted the residual background level in each column by measuring the median of pixels that were at least 5 pixels away from the center of any fiber. Finally, for each fiber, we used optimal extraction \citep{Horne1986} to sum the flux using the positions and standard deviations defined by the 1D Gaussian LSF profiles calculated from spectra of the telluric star.

Since the transmission of the telescope and instrument varies with wavelength due to the blaze function, we observed the M giant star \NMstar\ to determine the wavelength solution from the stellar and telluric spectral lines. This M0.5III star was chosen because it has myriad narrow stellar lines in the K~band and a well-known RV, making it a good target for wavelength reference. First, we modeled the wavelength solution as a spline using six nodes per order. \NMstar\ was modeled with a PHOENIX stellar spectrum \citep{Husser2013} assuming a temperature of $3800$~K, a surface gravity of $\mathrm{log}~g=1.5$, solar metallicity, and a fixed known RV of $-85.391$~km~s$^{-1}$ taken from SIMBAD\footnote{SIMBAD: \url{http://simbad.u-strasbg.fr/simbad/}}. The telluric transmission of Earth's atmosphere was modeled from a planetary spectrum generator model \citep{Villanueva2018}. To model the spectrum-dependent transmission, we then jointly fit for the wavelength solution, telluric parameters, and instrument and telescope transmission using the Nelder–Mead optimization \citep{Virtanen2020}. For more details on the wavelength calibration, we refer to \cite{Wang2021a}.

\begin{figure*}
    \centering
    \includegraphics[width=1\textwidth]{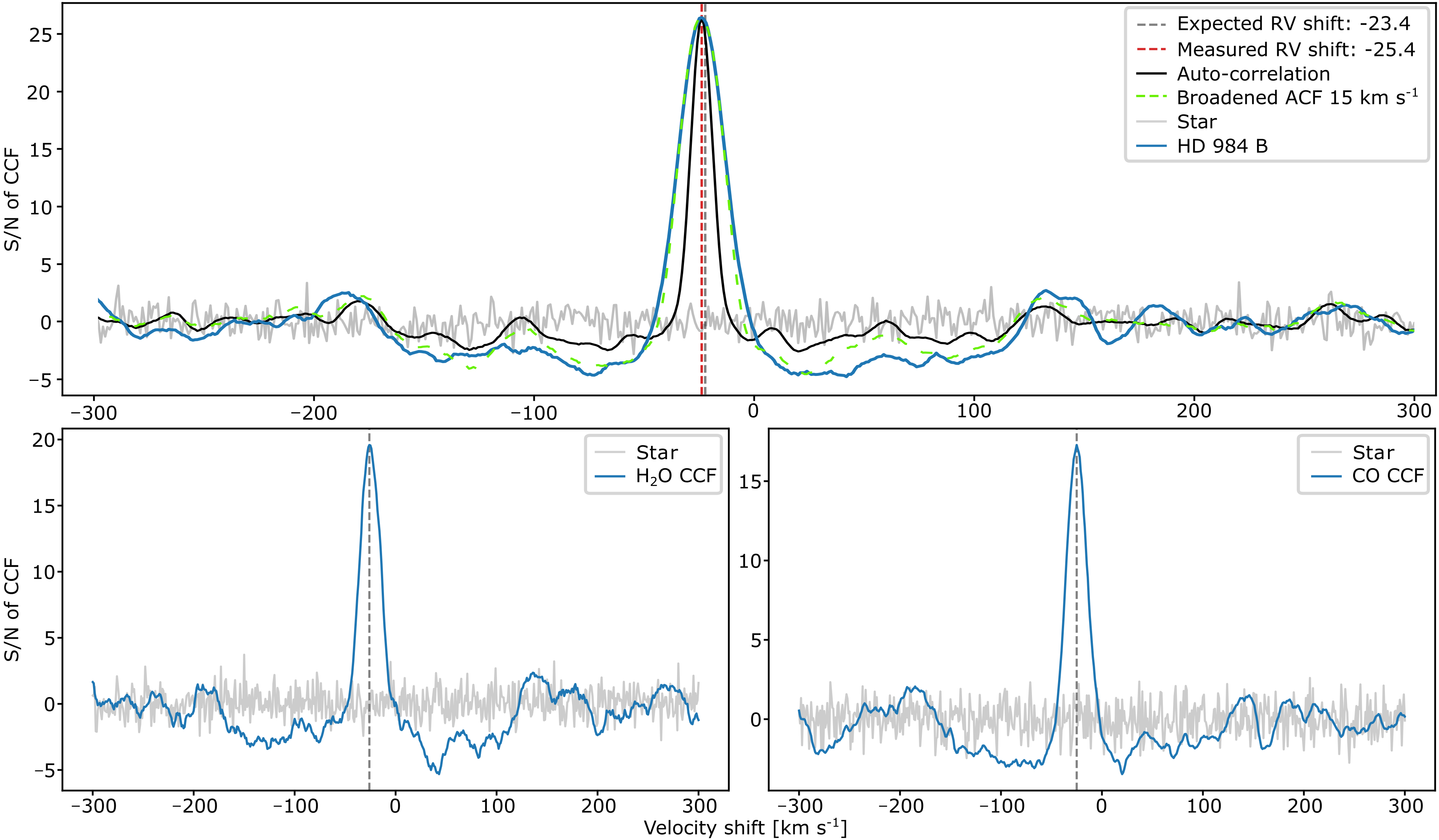}
    \caption{\label{fig:CCF} CCFs between the KPIC data (using the three orders from 2.29 to 2.49~$\mu$m) and different models. Top panel: CCFs between the KPIC data and a BT-Settl model (see Section~\ref{Preliminary_Analysis}). The CCF for \Ncomp\ is plotted in blue, and for comparison, we add the CCF of the stellar data in gray. The autocorrelation of the planetary model is also shown in black. The standard deviation of the wings of the CCF were used to estimate its noise and to normalize the CCF. The dashed vertical gray and red lines show the expected and measured RV of the companion ($-23.4~\mathrm{km~s^{-1}}$ and $-25.8~\mathrm{km~s^{-1}}$, respectively). Bottom panels: CCFs showing the H$_2$O and CO detection from the same KPIC data after using single-molecule templates.}
\end{figure*}

\subsection{Preliminary analysis}\label{Preliminary_Analysis}
Our extracted spectra of \Ncomp\ consist of a mixture of light coming from the brown dwarf companion and stellar speckles. We therefore used a forward model for the signal from the planet fiber ($\mathrm{D_p}$) via
\begin{ceqn}
\begin{equation}\label{Companion Data}
   \mathrm{D_p(\lambda)} = \mathrm{\alpha_p(\lambda)T(\lambda)P_{LSF}(\lambda) + \alpha_s(\lambda)T(\lambda)S_{LSF}(\lambda) + n(\lambda)}~,
\end{equation}
\end{ceqn}
where $\mathrm{T}$ is the transmission of the optical system (i.e., atmosphere, telescope, and instrument), and $\mathrm{P_{LSF}}$ and $\mathrm{S_{LSF}}$ are the spectrum from the planet and from the star, respectively, after convolution by the instrumental LSF. $\mathrm{\alpha_p}$ and $\mathrm{\alpha_s}$ are the scaling factor for the planetary and stellar brightness, respectively, and $\mathrm{n}$ is the noise. Because of the high S/N per spectral channel of the stellar spectra ($\sim 300$), the noise $\mathrm{n}$ was assumed to be negligible.

We explain next how we measured each parameter of Equation~\ref{Companion Data}. First, we measured the transmission $\mathrm{T}$ using a PHOENIX model mimicking the \NTstar\ spectrum assuming an effective temperature of 9600~K, a surface gravity of $\mathrm{log(g)}=3.5$, and solar metallicity. \NTstar\ was chosen because it is an A0V star, and thus, the created model has nearly no spectral lines in the K~band. This mitigates any errors due to an imperfect stellar spectrum. Then, $\mathrm{T}$ was obtained by dividing the observed telluric standard star data ($\mathrm{D_t}$) by the PHOENIX model of the star ($\mathrm{M_t}$),
\begin{ceqn}
\begin{equation}
   \mathrm{T} = \frac{\mathrm{D_t}}{\mathrm{M_t}}~.
\end{equation}
\end{ceqn}
To model the contribution of the stellar light that leaked into the planet fiber (i.e., $\mathrm{T \times S_{LSF}}$), we then used the on-axis observation of the star \Nstar\ itself. Finally, $\mathrm{P_{LSF}}$ was determined using a planetary atmospheric model, and the variables $\mathrm{\alpha_p}$ and $\mathrm{\alpha_s}$ were measured by finding the best fit to the data.

As a preliminary step, we confirmed the detection of \Ncomp\ by applying a modified cross-correlation analysis, as described in \cite{Ruffio2019} and \cite{Wang2021a}. This technique consists of estimating the maximum likelihood value for both the planet and star flux as a function of RV shift for a given planet template. More explicitly, we searched for the maximum likelihood between our equation~(\ref{Companion Data}) and our observed planetary data. To do this, we built a companion model and a stellar model in order to fit our data. Since the S/N of KPIC is optimized for wavelengths around 2.3~$\mu$m (i.e., where the CO has a series of strong absorption lines), we focused our analysis on three spectral orders, from 2.29 to 2.49~$\mu$m. These three orders are the best suited for an analysis of \Ncomp\ because they contain the strongest absorption lines from the companion and have few telluric absorption lines. 

Our companion model was first built using a BT-Settl-CIFIST model grid \citep{Baraffe2015}. A BT-Settl planetary atmospheric model was chosen as it was the only publicly available grid of models with a spectral resolution $\mathrm{R}>35,000$ and with an effective temperature $\mathrm{T_{eff}}\sim2800$~K that includes clouds. The presence of clouds in our model might be particularly important for high-resolution spectra because clouds change the depths of molecular absorption lines \citep{Hood2020, Molliere2020}. Our planetary model was then shifted to fit for the RV. Then, the model was rotationally broadened by a projected rotation rate $v\sin i$ and convolved using the instrumental LSF. Finally, the companion model was multiplied by the telluric response function $\mathrm{T}$. In the last steps, we removed the continuum variations of the models and planetary data. A high-pass filtering was applied with an optimal median filter size of 100 pixels ($\sim 0.002~\mu$m), as determined in \cite{Xuan2022}. After scaling the companion and speckle models using different $\mathrm{\alpha_p}$ and $\mathrm{\alpha_s}$ for each order, we fit the total model to the data and searched for the maximum likelihood value. 

From this modified and normalized cross-correlation between the KPIC data and the BT-Settl model, we confirmed the clear detection of \Ncomp\ with an S/N of $\sim25$ (see Figure~\ref{fig:CCF}). Additionally, we find that H$_2$O and CO are detected with an S/N of 20 and 18, respectively, using single-molecule templates (built from the best parameters found from our \pRT\ retrievals; see Section~\ref{results}). We note that while several other atomic lines are expected in the stellar atmospheres of targets with $\mathrm{T_{eff} = 2700~K}$ in the K band, such as Na, Ca, Al and Fe, they only appear as a handful of single lines, which makes it hard to properly identify them with CCFs. The standard deviation of the wings of the CCF were used to estimate its noise and to normalize the CCF. We measured a RV of $-25.8~\mathrm{km~s^{-1}}$, which is within the $1\sigma$ error bar from the expected value ($-23.4 \pm 5~\mathrm{km~s^{-1}}$) obtained from the orbital prediction tool \texttt{whereistheplanet} (based on the GRAVITY data, unpublished), by taking the RV of its host star into account and adding the barycentric correction, including the effect caused by the Earth's rotation. Finally, using the atmospheric template, we broadened the results of its autocorrelation function and found that the best fit to the CCF was approximately $15~\mathrm{km~s^{-1}}$, giving us a first estimate of the rotational velocity of \Ncomp. This result is also shown in Figure~\ref{fig:CCF}, where we plot the ACF and its broadened version, both rescaled to the peak value of the CCF.


\section{Spectral analysis}\label{Spectral_Analysis}
\subsection{Atmosphere retrieval}
After the preliminary analysis of our data, which confirmed the presence of the planetary companion using a BT-Settl model,  we repeated our analysis using a free retrieval framework based on the open-source radiative transfer code \pRT\ \citep{Molliere2019}. This section describes how we built our models to fit the data.

\subsubsection{Opacities}
Following our preliminary analysis (described in Sec~\ref{Preliminary_Analysis}), our goal was to combine a companion model and a stellar speckle model to fit our data. Instead of using self-consistent grid models, \pRT\ allows more flexibility to fit the data, which provides much more detailed information about the atmospheric properties. This method has been used in previous studies \citep[e.g.,][]{Molliere2020, Xuan2022} that proved that it can retrieve the chemical abundances, vertical temperature structure, and cloud properties of exoplanets and brown dwarfs. The flexibility of the retrieval approaches makes them ideal for exploring the properties of complex objects such as brown dwarfs and exoplanets, but it can also lead to occasional unphysical results. This is why the choice of our priors and boundaries is important (see Table~\ref{tab:priors}).

In order to correctly set our model up, we first used the line-by-line opacity sampling method in \pRT\ for the retrieval, including opacities for H$_2$O, $^{12}$CO, $^{13}$CO, CO$_2$, FeH, NH$_3$, and H$_2$S. The choice of these opacities was made based on the relatively high effective temperature of \Ncomp\ (2730~K) and on the selected observed wavelength range in the K~band (between 2.29 and 2.49~$\mu$m). For completeness in our retrieval, we also searched for methane in the atmosphere of \Ncomp. We used the HITEMP~CH$_4$ line list from \cite{Hargreaves2020}, which was converted into opacities for use in \pRT\ as explained in \citet{Xuan2022}. Collision-induced absorption (CIA) opacities of H2-H2, H2-He, and H- were also accounted for in our retrieval as these species are important absorbers in the K~band.

\subsubsection{Temperature model}
We set the vertical extent of the atmosphere of our brown dwarf using a pressure-temperature ($\mathrm{P-T}$) profile between $\mathrm{P=10^{-4} - 10^{2}}$ bars and followed the $\mathrm{P-T}$ profile definition from \cite{Molliere2020}. This definition parameterizes the vertical temperature profile of the atmosphere into three different parts: high altitudes, the photosphere (middle altitudes), and the troposphere (low altitudes), where the spatial coordinate of the temperature model is an optical depth ($\tau$), defined via
\begin{ceqn}
\begin{equation}\label{tau}
   \tau = \delta\mathrm{P}^\alpha~,
\end{equation}
\end{ceqn}
with $\delta$ and $\alpha$ being two free parameters. It should be noted that $\alpha$ was restricted to vary only between 1 and 2, following \cite{Robinson2012}. 

For the high altitudes (between the top of the atmosphere to $\tau=0.1$), the atmosphere was fit with three temperature points equidistant in log(P) space, which were treated as free parameters. These three temperature points were defined such that they were colder than the highest point of the photosphere and such that they monotonically decreased in temperature with increasing altitude in order to prevent the formation of temperature inversions. For the photosphere (between $\tau=0.1$ to the radiative-convective boundary), the temperature was set according to the Eddington approximation with $\mathrm{T}_0$ as the internal temperature. Finally, for the troposphere (between the radiative-convective boundary to the bottom of the atmosphere), it was forced to follow the moist adiabatic temperature gradient as soon as the atmosphere was found to be Schwarzschild-unstable \citep{Molliere2020}.

\subsubsection{Chemistry model}
We followed the equilibrium chemistry model described in \cite{Molliere2019} to compute the abundance of each absorbing molecule. These abundances were defined as a function of pressure, temperature, carbon-to-oxygen number ratio (C/O) and metallicity ([Fe/H]). The temperature ranged from 500 to 4000~K, the C/O values from 0.1 to 1.6, and the [Fe/H] from -1.5 to 1.5. Since the S/N of our data is relatively high, we also studied the abundance of the $\mathrm{^{13}CO}$ isotope by fitting in our retrieval the $\mathrm{^{12}CO/^{13}CO}$ ratio (ranging from 0 to 6 in log-scale).

To account for the possibility of disequilibrium chemistry, we included a quench pressure term ($\mathrm{P_{quench}}$) as a free parameter. This disequilibrium chemistry can arise when the atmospheric mixing timescale is shorter than the chemical reaction timescale. For atmospheric pressures $\mathrm{P}<\mathrm{P_{quench}}$, we therefore fixed the abundances of $\mathrm{H_2O}$, $\mathrm{CO}$, and $\mathrm{CH_4}$ using the equilibrium value found at $\mathrm{P_{quench}}$, following the results from \cite{Zahnle2014}.

\begin{table*}
    \centering
    \caption{Priors for the \Ncomp\ retrieval.}
    \begin{tabular}{ll|ll}
    \hline\hline
    \rule{0pt}{2.5ex}Parameter  &       Prior & Parameter       &       Prior\\[.5ex]
    \hline
    \rule{0pt}{2.5ex}RV ($\mathrm{km~s}^{-1}$) & $\mathcal{U}$(-40, -10) & $\mathrm{T_{int}}$ (K) & $\mathcal{U}$(500, 4000) \\
    $v\sin i$ ($\mathrm{km~s}^{-1}$) & $\mathcal{U}$(5, 25) & $\mathrm{T_3}$ (K) & $\mathcal{U}$(0, $\mathrm{T_{connect}~^{(b)}}$) \\
    $\mathrm{R_{comp}}$ ($\mathrm{R_{Jup}}$) & $\mathcal{U}$(0.5, 2.3) & $\mathrm{T_2}$ (K) & $\mathcal{U}$(0, T3) \\
    $\mathrm{\alpha_p}$ ($\%$) & $\mathcal{U}$(30, 110) & $\mathrm{T_1}$ (K) & $\mathcal{U}$(0, T2) \\
    $\mathrm{\alpha_s}$ ($\%$) & $\mathcal{U}$(-10, 70) & $\mathrm{log(P_{quench})}$ & $\mathcal{U}$(-4, 3) \\
    $\mathrm{\sigma_s}$ & $\mathcal{U}$(1.0, 1.2) & $\alpha$ & $\mathcal{U}$(1, 2) \\
    log g & $\mathcal{U}$(3.5, 6) & C/O & $\mathcal{U}$(0.1, 1.6) \\
    log($\mathrm{^{12}CO/^{13}CO}$) & $\mathcal{U}$(0, 6) & $[$Fe/H$]$ & $\mathcal{U}$(-1.5, 1.5) \\
    $\mathrm{log(\delta)}$ & $\mathrm{P_{phot}\in[10^{-3}, 10^{2}]~^{(a)}}$ & & \\[.5ex]
    \hline
    \multicolumn{4}{c}{\rule{0pt}{3.5ex}Additional parameter for gray model}\\[.5ex]
    \hline
    \rule{0pt}{2.5ex}log(gray opacity) ($\mathrm{cm^{2}~g^{-1}}$) & $\mathcal{U}$(-6, 1) & &\\[.5ex]
    \hline
    \multicolumn{4}{c}{\rule{0pt}{3.5ex}Additional parameters for cloudy model}\\[.5ex]
    \hline
    \rule{0pt}{2.5ex}$\mathrm{log(K_{zz}~/~cm^2~s^{-1})}$ & $\mathcal{U}$(5, 13) & $\mathrm{f_{sed}}$ & $\mathcal{U}$(0, 10) \\
    $\mathrm{\sigma_g}$ & $\mathcal{U}$(1.05, 3) & $\mathrm{log(\Tilde{X}_{C})~^{(c)}}$ & $\mathcal{U}$(-2.3, 1) \\[.5ex]
    \hline
    \end{tabular}
    \tablefoot{For all parameters, we adopted uniform or log-uniform priors (shown as $\mathcal{U}$) where the lower and upper boundaries are represented. $\mathrm{^{(a)}}~\mathrm{P_{phot}}$ is the pressure for $\tau=1$ and $\mathrm{^{(b)}}~\mathrm{T_{connect}}$ is the uppermost temperature of the photospheric layer, calculated by setting $\tau=0.1$ in the Eddington approximation (see \citealt{Molliere2020}). The priors for T$_3$, T$_2$, and T$_1$ were set to ensure a temperature profile that decreased monotonically with altitude to avoid temperature inversions. $\mathrm{^{(c)}~log(\Tilde{X}_{C})}$ represents the scaling factor for the cloud mass fraction for each of the cloud species used in our retrieval (i.e., $\mathrm{Al_2O_3}$, Fe, and $\mathrm{MgSiO_3}$), defined such that $\mathrm{log(\Tilde{X}_{C})}=0$ refers to a fraction equal to the equilibrium mass fraction.} \label{tab:priors}
\end{table*}

\subsubsection{Clouds}\label{4.Clouds}
In our analysis, we considered both clear (i.e., without clouds) and cloudy models to compare our retrieved abundances. For the latter, we used the EddySed model from \cite{Ackerman2001} as implemented in \pRT, where the cloud particles both absorb and scatter the absorbing photon from the atmosphere. By taking the effective temperature of \Ncomp\ into account and comparing the results from \cite{Wakeford2017} and \cite{Gao2020} for the most likely cloud species as a function of temperature, we decided to consider three different cloud species: Fe, $\mathrm{MgSiO_3}$, and $\mathrm{Al_2O_3}$. We assumed their crystalline phases. In particular, $\mathrm{Al_2O_3}$ appears to be the most important species of the three when we compare their condensation curves with different thermal profiles from the Sonora atmospheric model \citep{Marley2021}. In addition, $\mathrm{MgSiO_3}$ is also expected to be one of the most important cloud species for substellar objects with $\mathrm{T_{eff} > 950~K}$ because its nucleation energy barriers are low and the elemental abundances of Mg, Si, and O are relatively high \citep{Gao2020}.

Three free parameters ($\mathrm{f_{sed}}$, $\mathrm{K_{zz}}$, and $\mathrm{\sigma_g}$) are needed to control the mean particle size, cloud mass fraction, and particle size distribution independently \citep{Ackerman2001}. First, by setting $\mathrm{f_{sed}}$ we can retrieve the sedimentation efficiency. Then, using the vertical eddy diffusion coefficient ($\mathrm{K_{zz}}$), we set the particle size for a given $\mathrm{f_{sed}}$ value. Finally, by fitting for the width of the log-normal size distribution ($\mathrm{\sigma_g}$), we specified the particle size distribution. Following \cite{Molliere2020}, we let these parameters vary independently between 0 to 10 for $\mathrm{f_{sed}}$, 5 to 13 for $\mathrm{log(K_{zz})}$ (with $\mathrm{K_{zz}}$ expressed in $\mathrm{cm^2~s^{-1}}$), and 1.05 to 3 for $\mathrm{\sigma_g}$. We also took into account as free parameters the scaling factor for the cloud mass fraction at the cloud base, $\mathrm{log(\Tilde{X}_{C})}$, for each of the cloud species used in our retrieval (i.e., $\mathrm{Al_2O_3}$, Fe, and $\mathrm{MgSiO_3}$), defined such that $\mathrm{log(\Tilde{X}_{C})}=0$ refers to the mass fraction predicted for the cloud species when equilibrium condensation is assumed at the cloud base location. 

\begin{figure*}
    \centering
    \includegraphics[width=1\textwidth]{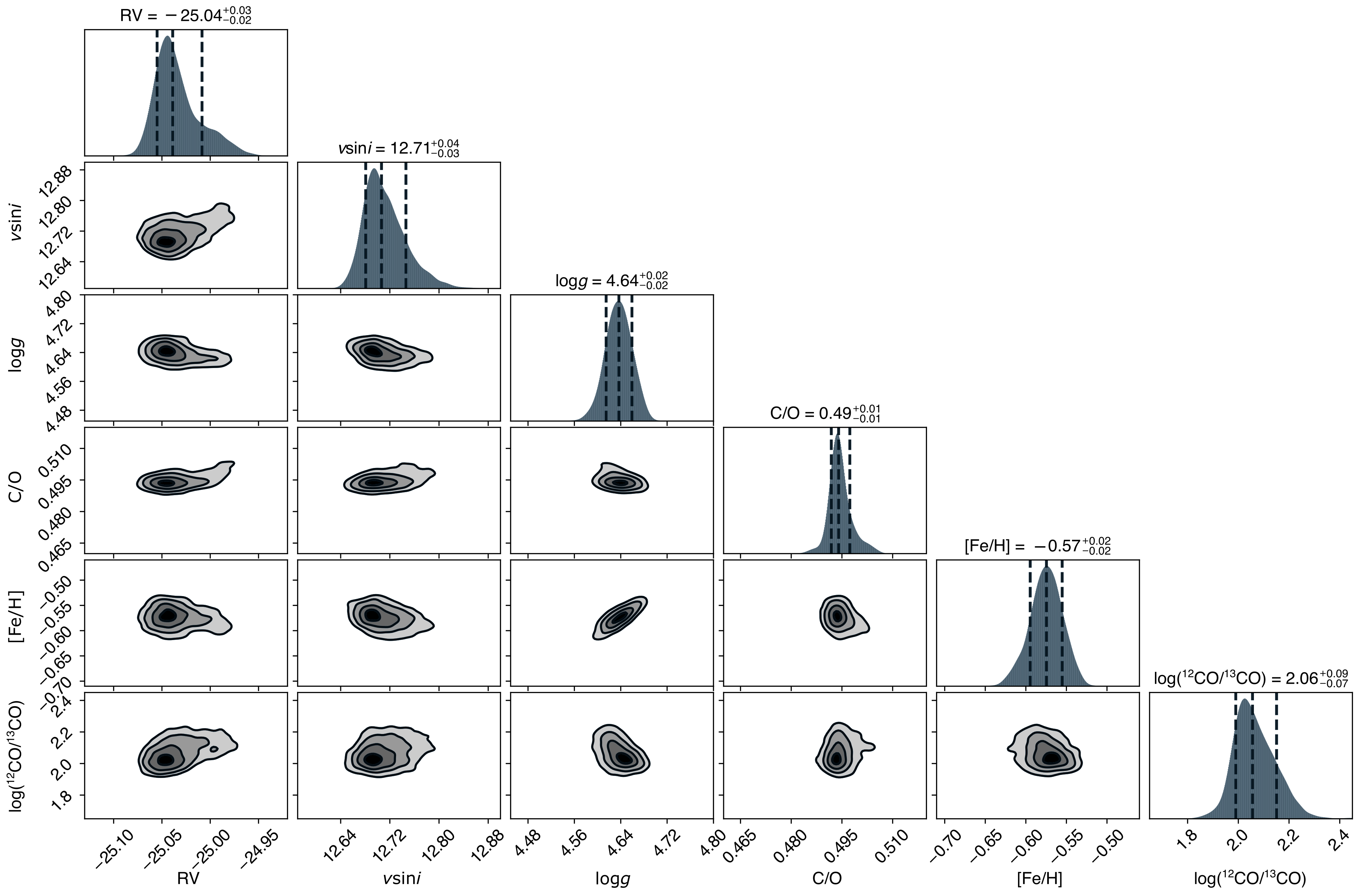}
    \caption{\label{fig:Corner_1} Posterior distributions for some of the key parameters of our clear model retrieval for \Ncomp. The titles at the top of each histogram show the median and the 1$\sigma$ error.}
\end{figure*}

\subsection{Fitting with nested sampling}\label{results}
\subsubsection{Priors}
To derive the posterior distributions for \Ncomp\, we used the nested-sampling approach as implemented in \texttt{dynesty} \citep{Speagle2020}. We chose nested sampling (over the more classical Markov chain Monte Carlo, MCMC) because it is faster and can approximate the probability of the model via the Bayesian inference, which facilitates the model comparison. Specifically, we used the dynamic nested-sampling method, which allocates live points dynamically instead of using a constant number of live points, to improve the posterior density estimate. We used 500 live points to start the initial run, and we adopted the stopping criterion that the estimated contribution of the remaining prior volume to the total evidence be less than $1\%$. We also repeated our retrieval using 1000~initial live points and found that the evidence remained the same.

Similar to \cite{Molliere2020}, we adopted uniform or log-uniform priors for all parameters. Table~\ref{tab:priors} presents the parameters and their priors as we used them in our retrieval. We distinguished between those needed for a clear model and the additional parameters needed for a cloudy model. The $\mathrm{P_{phot}}$ quantity was used to constrain the $\delta$ parameter from Eq.~\ref{tau}, and it is defined as the pressure for $\tau=1$. $\mathrm{T_{connect}}$ is the uppermost temperature of the photospheric layer, calculated by setting $\tau=0.1$ in the Eddington approximation (see \citealt{Molliere2020}), and it was used to constrain the T$_3$ parameter. We additionally fit the radius of \Ncomp. Using evolutionary and mass-radius models \citep{Baraffe2003, Baraffe2008}, we decided to constrain the radius between 0.5 to 2.3~$\mathrm{R_{Jup}}$. We chose the priors of the RV and of the $v\sin i$ parameters based on the results of our preliminary analysis (see Section~\ref{Preliminary_Analysis}). Finally, we also took the uncertainty in the line spread function into account (noted $\mathrm{\sigma_s}$). This uncertainty is due to the difference of a factor of 1.13 in the focal lengths in the spatial and dispersion directions of the NIRSPEC spectrograph \citep{Robichaud1998}.  Following \cite{Wang2021a}, we let $\mathrm{\sigma_s}$ vary between 1.0 and 1.2 to conservatively account for any systematics.

\begin{figure*}
    \centering
    \includegraphics[width=1\textwidth]{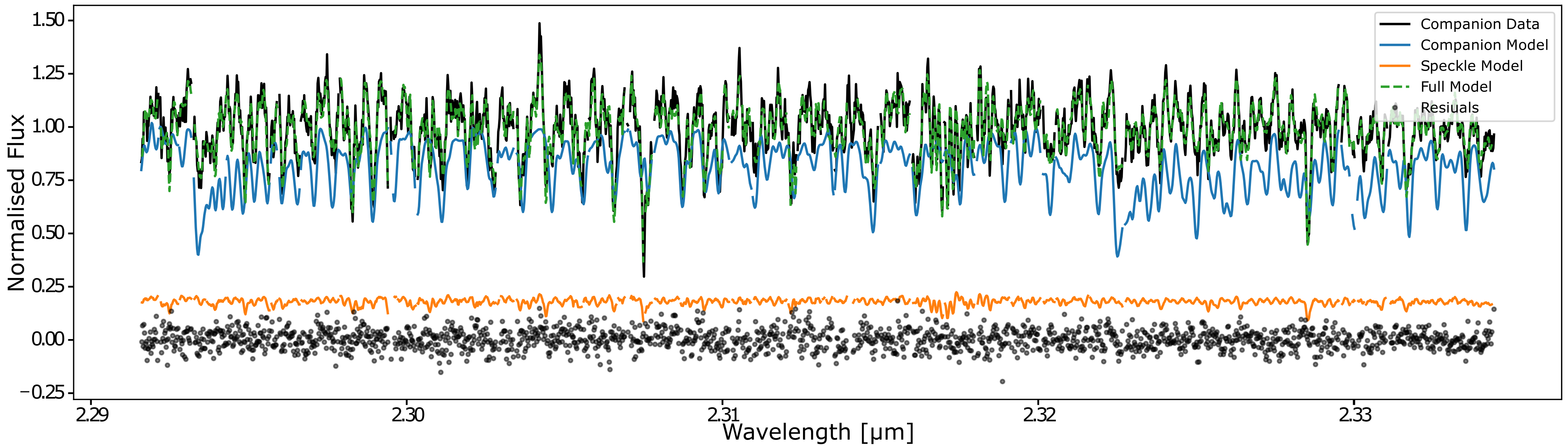}
    \caption{\label{fig:Models} Retrieval results normalized for \Ncomp. The KPIC data are plotted in black. The companion model retrieved using nested sampling and \pRT\ is plotted in blue, scaled using the retrieved $\mathrm{\alpha_{p}}$ from our fit. The companion model does not include tellurics (i.e., we show the model before it was multiplied to the telluric response function) to focus on the molecular features, but tellurics are included in our fits. The stellar model used to model the speckle contribution is plotted in orange, scaled using the retrieved $\mathrm{\alpha_{s}}$ from our fit. The full model (i.e., combination of the companion and speckle model) is shown in dashed green. The residuals between the KPIC data and our retrieved model are shown as the gray points to the bottom. The x-axis is restricted for clarity, and we focus on one order only instead of the full data range (between 2.29 and 2.49~$\mu$m).}
\end{figure*}

\begin{table*}
    \centering
    \caption{Comparison of the spectrum retrievals for \Ncomp}
    \begin{tabular}{l|ccc|c}
    \hline\hline
    \rule{0pt}{2.5ex}Model      &       C/O & $[$Fe/H$]$ & $\mathrm{log(^{12}CO/^{13}CO})$      &       ln(B) \\[.5ex]
    \hline
     \rule{0pt}{2.5ex}Clear & $0.49^{+0.01}_{-0.01}$ & $-0.57^{+0.02}_{-0.02}$ & $2.06^{+0.09}_{-0.07}$ & $0$ \\ [0.2cm]
     Gray opacity cloud & $0.50^{+0.01}_{-0.01}$ & $-0.62^{+0.02}_{-0.02}$ & $2.03^{+0.07}_{-0.08}$ & $-0.6$\\ [0.2cm]
     EddySed ($\mathrm{Al_2O_3}$, cd) & $0.49^{+0.01}_{-0.01}$ & $-0.55^{+0.02}_{-0.02}$ & $2.04^{+0.07}_{-0.06}$ & $1.0$\\ [0.2cm]
     EddySed ($\mathrm{Al_2O_3 + MgSiO_3}$, cd) & $0.52^{+0.01}_{-0.01}$ & $-0.47^{+0.07}_{-0.07}$ & $1.92^{+0.03}_{-0.06}$ & $-2.2$ \\ [0.2cm]
     EddySed ($\mathrm{MgSiO_3 + Fe}$, cd) & $0.50^{+0.01}_{-0.01}$ & $-0.62^{+0.02}_{-0.02}$ & $1.99^{+0.08}_{-0.13}$ & $3.3$ \\ [0.2cm]
     EddySed ($\mathrm{Al_2O_3 + MgSiO_3 + Fe}$, cd) & $0.50^{+0.01}_{-0.01}$ & $-0.63^{+0.03}_{-0.02}$ & $1.97^{+0.19}_{-0.17}$ & $2.2$ \\[.5ex]
    \hline
    \multicolumn{5}{c}{\rule{0pt}{3.5ex}Additional retrieval without $^{13}$CO}\\[.5ex]
    \hline
    \rule{0pt}{2.5ex}Clear without $^{13}$CO & $0.49^{+0.01}_{-0.01}$ & $-0.56^{+0.02}_{-0.02}$ & $-$ & $-7.8$\\ [0.2cm]
    EddySed ($\mathrm{MgSiO_3 + Fe}$, cd) without $^{13}$CO & $0.49^{+0.01}_{-0.01}$ & $-0.56^{+0.02}_{-0.01}$ & $-$ & $-7.1$ \\ [0.2cm]
    \hline
    \end{tabular}
    \tablefoot{Comparison of some of the retrieved parameters obtained using clear and cloudy models. The rightmost column lists the measured Bayes factor for each retrieval, with the clear model as the baseline model with $\mathrm{ln(B) = 0}$. We adopted the EddySed ($\mathrm{MgSiO_3 + Fe}$, cd) model as our best-fit model since it shows the highest Bayes factor.} \label{tab:evidence}
\end{table*}

\subsubsection{Retrieval with a clear model}\label{clear model}
Using a clear model, we were able to precisely retrieve most of the companion parameters. For clarity, a corner plot presenting selected retrieved parameters is shown in Figure~\ref{fig:Corner_1}. An extended corner plot can be found in the appendix (see Figure~\ref{fig:Corner_2}). For instance, we retrieved an RV of $-25.04^{+0.03}_{-0.02}$~km~s$^{-1}$ and a $v\sin i$ of $12.71^{+0.04}_{-0.03}$~km~s$^{-1}$, which correspond well to our first guess from our CCF (see Figure~\ref{fig:CCF}) and is comparable to the rotation rates observed for brown dwarfs with similar spectral types \citep{Konopacky2012}. We also retrieved a C/O ratio of $0.49 \pm 0.01$ for \Ncomp\, which is consistent with the solar value and is $<1 \sigma$ consistent with that of its host star (due to its large error bar, $0.40 \pm 0.20$). Using our clear model, we were also able to detect the $^{13}$CO isotopolog in the atmosphere of \Ncomp\, and we measure an isotopolog $\mathrm{^{12}CO/^{13}CO}$ ratio of $115^{+26}_{-17}$. 

Figure~\ref{fig:Models} shows the results of our analysis from the parameters retrieved using nested sampling and \pRT. We plot the best-fit companion model, the speckle model, and the full model, which is the sum of the previous models. We also plot the residuals, for which no evidence of correlated noise or strong systematics was found. While we precisely retrieved some parameters with our clear model, other parameters were not well constrained (see Figure~\ref{fig:Corner_2}). Our retrieval failed, for example, to properly constrain the radius of \Ncomp\ or the quench pressure term ($\mathrm{P_{quench}}$). This outcome can be explained by the hot temperature of \Ncomp, which makes it extremely difficult to detect $\mathrm{CH_4}$ and therefore precludes a robust constraint on $\mathrm{P_{quench}}$.

\subsubsection{Comparison with cloudy models}\label{cloudy model}
After fitting a clear model, we investigated different types of cloudy models. For instance, we studied a gray cloud model that adds a constant cloud opacity to the atmosphere to examine the possibility of finding clouds at lower pressures. We also tested a variety of combinations for the cloud species using $\mathrm{Al_2O_3}$, Fe, and $\mathrm{MgSiO_3}$. For each retrieval, we measured and compared their Bayes factor (B) in order to assess the probability of each model. 
Table~\ref{tab:evidence} summarizes the different tests showing the models used, some of the retrieved parameters obtained, and the measured Bayes factor for each test compared to the clear model.

From these results, we first observe that all models are consistent with each other, with Bayes factor values that do not vary too much relative to the clear model. We also note that despite the high temperature of \Ncomp, $\mathrm{Al_2O_3}$ does not have a major impact in its atmosphere, as might have been expected (see Section~\ref{4.Clouds}). This result could be due to the fact that $\mathrm{Al_2O_3}$ has a much lower relative cloud mass than Fe and $\mathrm{MgSiO_3}$ (by a factor $10-20$; see \citealt{Wakeford2017}). Given their similar Bayes factor, these results also show that clouds do not seem to play an important role in the high dispersion spectroscopy of the atmosphere of \Ncomp in general. This result could be explained by the fact that the cloudy parameters of our retrieval mostly affect the continuum, which is ignored in the analysis of our KPIC data. We stress, however, that for the remainder of the paper, we consider the values retrieved from the EddySed ($\mathrm{MgSiO_{3} + Fe}$, cd) model as our most likely results. This model shows the highest Bayes factor, with $\mathrm{ln(B) = 3.3}$, which corresponds to a $\sim4.3\sigma$ preference for the \cite{Trotta2008} scale (see Figure~\ref{fig:Corner_3} for the posterior distribution for our best-fit model).

Table~\ref{tab:evidence} also shows that the isotopolog $^{13}$CO is detected in the atmosphere of \Ncomp\ in each retrieval, and the values obtained are consistent among the models, with an isotopolog $\mathrm{^{12}CO/^{13}CO}$ ratio of $98^{+26}_{-19}$. To confirm this detection, we also tested the retrieval using a clear model without $^{13}$CO and an EddySed ($\mathrm{MgSiO_3 + Fe}$, cd) retrieval without $^{13}$CO models (see the bottom of Table~\ref{tab:evidence}). Bayes factors of $\mathrm{ln(B) = -7.8}$ and $-7.1$ were measured, respectively, corresponding to a $\sim6.3\sigma$ and a $\sim6.1\sigma$ using the \cite{Trotta2008} scale. This confirms the detection of $^{13}$CO at a significance of $\sim6\sigma$ sigma.

\begin{figure}
    \centering
    \includegraphics[width=\columnwidth]{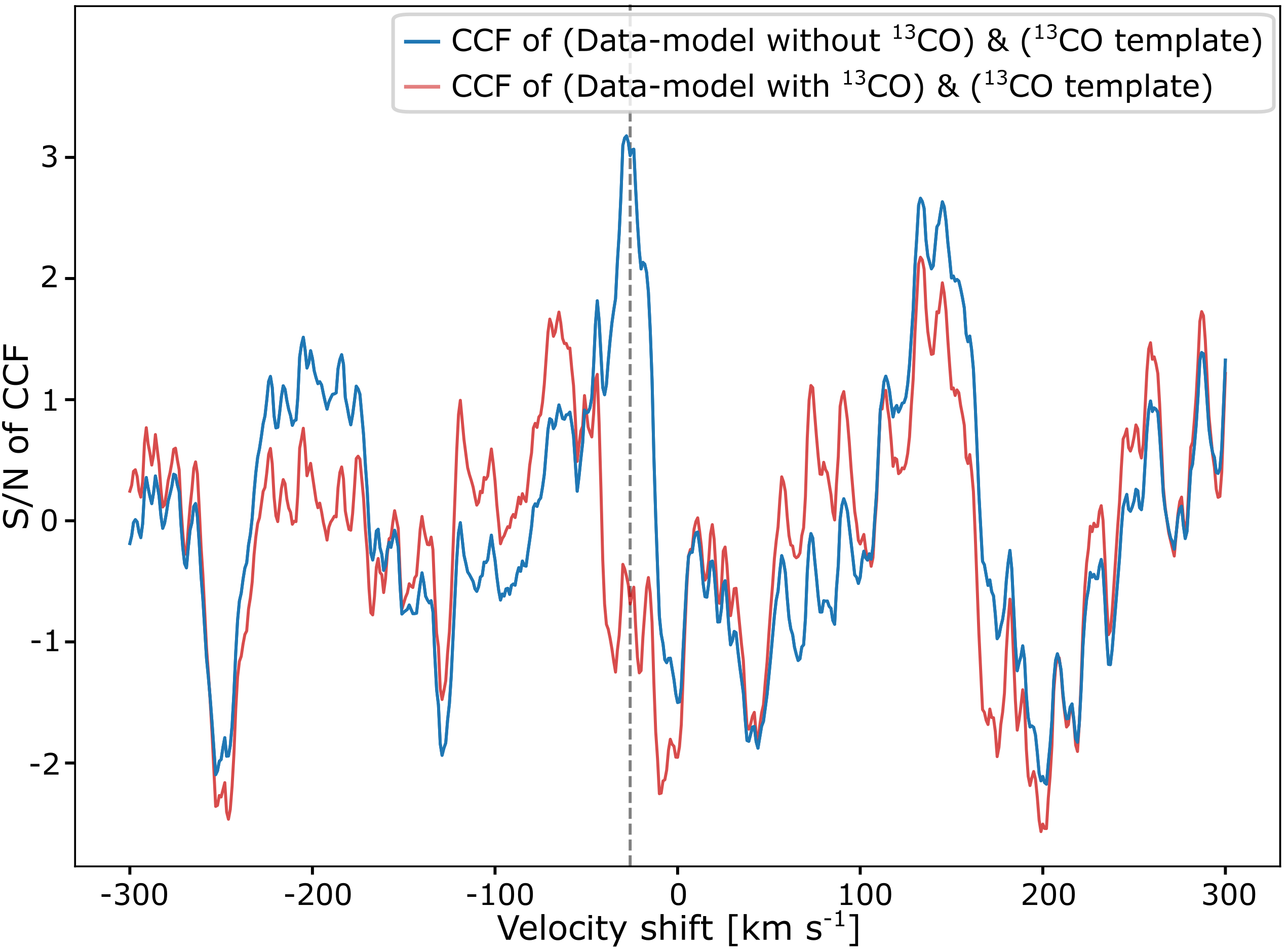}
    \caption{\label{fig:13COCCF} CCFs between residuals of our best-fit full model and a $^{13}$CO template. The CCF shown in blue corresponds to the residuals between our KPIC data and our full model without taking $^{13}$CO into account. Conversely, the CCF in red corresponds to the residuals between our KPIC data and our full model while taking $^{13}$CO into account. The peak in the blue CCF (that is absent in the red CCF) at the exact RV shift measured for \Ncomp\, shown with a vertical gray line, confirms the presence of $^{13}$CO in our data.}
\end{figure}

To prove that this detection is not due to our retrieval simply fitting some artifacts, we also used the cross-correlation approach as an alternative detection method. By using a CCF, we expect a clear peak for the species at the companion position. However, it is important to note that in contrast to what was done for $\mathrm{H_2O}$ and CO (see Figure~\ref{fig:CCF}), we cannot simply cross-correlate our data with a $^{13}$CO template because a pure $^{13}$CO template interferes too strongly with other species (e.g., $\mathrm{H_2O}$ and CO), which could lead to a false detection. Instead, we used the residuals between our data and our best-fit full model without taking $^{13}$CO into account. Thus, if $^{13}$CO is present in our data, it should be highlighted in these residuals. Figure~\ref{fig:13COCCF} shows the CCF between these residuals and a $^{13}$CO template in blue. As expected, a peak appears at the exact RV shift measured for \Ncomp. This confirms the detection of $^{13}$CO. In comparison, we repeated the same analysis, but this time, took $^{13}$CO into account in our best-fit full model. The residuals obtained by this approach should therefore remove $^{13}$CO. From these second residuals, we indeed observe no peak (i.e., no detection) in the new CCF, plotted in red, at the companion position, as expected.

Table~\ref{tab:evidence} also shows that the measured C/O ratios all match that of the host star ($0.40 \pm 0.20$), with a C/O ratio of $0.50\pm0.01$. Similarly, the measured metallicity values are all consistent in the different models, and we infer that \Ncomp\ has a substellar metallicity with $[$Fe/H$] = -0.62^{+0.02}_{-0.02}$. Hence, we find that the C/O ratio, the metallicity, and the $^{13}$CO ratio values do not strongly rely on the accurate inference of the temperature structure and clouds (as was found in \citealt{Molliere2020, Burningham2021, Zhang2021a}), and that the values found for these parameters are independent of the assumptions on the clouds.

\begin{figure*}
    \centering
    \includegraphics[width=1\textwidth]{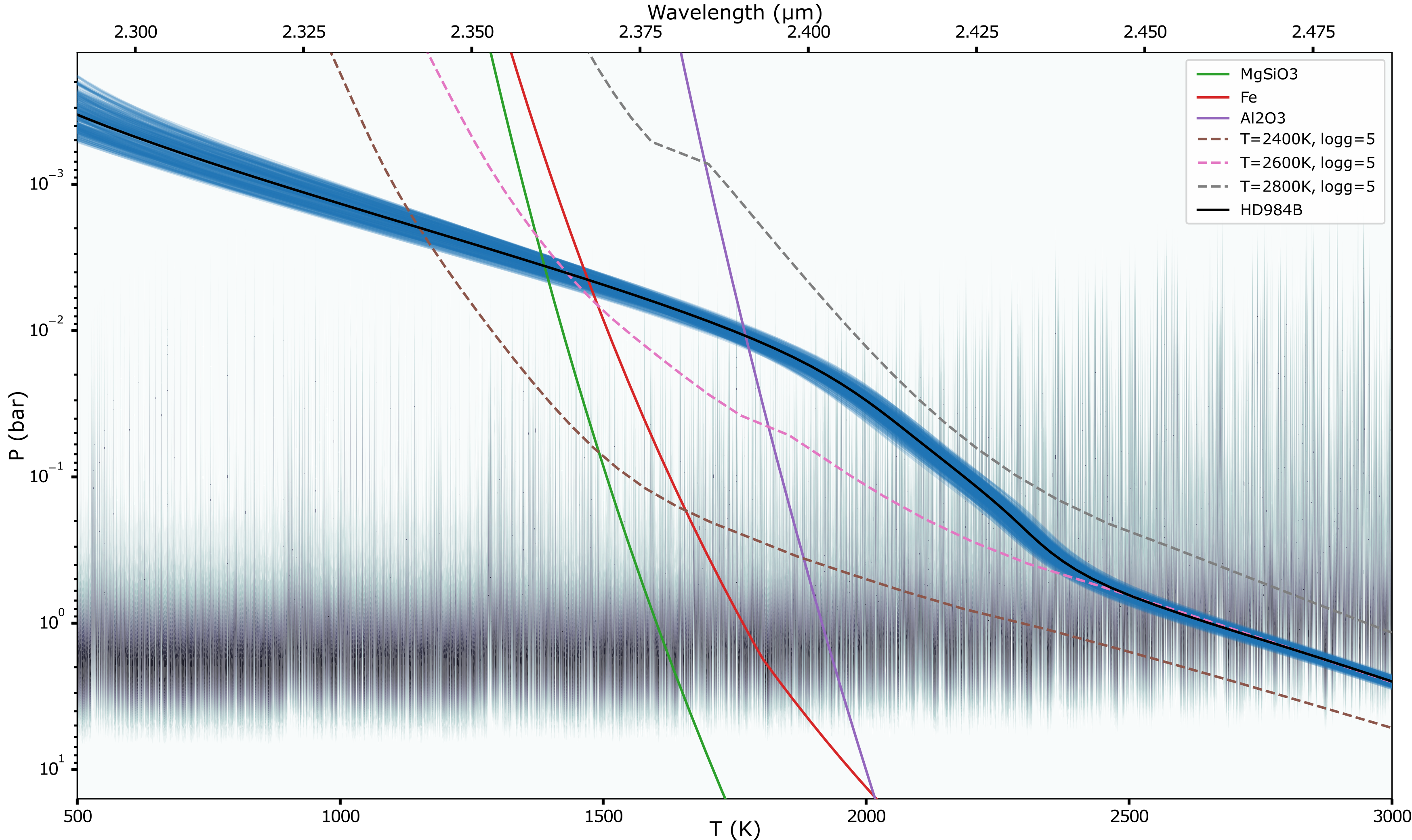}
    \caption{\label{fig:PT_profile} The $\mathrm{P-T}$ profile from our best-fit retrieval (model $\mathrm{MgSiO_{3} + Fe}$). The best-fit profile is shown in black, and 200 random draws from the posterior are shown in blue. The condensation curves for different cloud species ($\mathrm{MgSiO_{3}}$, Fe and $\mathrm{Al_{2}O_{3}}$) are also plotted. In addition, we show multiple Sonora $\mathrm{P-T}$ profiles as dashed lines \citep{Marley2021} with similar properties as \Ncomp. Finally, we plot the emission contribution function (in wavelength, top axis) as contours, which quantifies the relative importance of the emission in a given pressure layer to the total at a given wavelength \citep{Molliere2019}.}
\end{figure*}

Using the parameters retrieved from our best-fit model, we show in Figure~\ref{fig:PT_profile} the retrieved $\mathrm{P-T}$ profiles. For comparison, we also plot three cloud condensation curves (for $\mathrm{MgSiO_3}$, Fe, and $\mathrm{Al_2O_3}$) and three different Sonora $\mathrm{P-T}$ profiles \citealt{Marley2021} with properties similar to \Ncomp. These can be used to confirm where the cloud condensation curves are expected to intersect the $\mathrm{P-T}$ profile. As an example, the Sonora thermal profile of $\mathrm{T_{eff} = 2600~K}$ and $\mathrm{logg = 5.0}$ intersects the condensation curve of $\mathrm{Al_2O_3}$ near $0.04$~bar, in comparison with the condensation curves of Fe and $\mathrm{MgSiO_3}$, which intersect at lower pressure, around $0.007$ and $0.003$~bar. Finally, we added the emission contribution function to Figure~\ref{fig:PT_profile} to quantify the relative importance of the emission. This emission contribution function shows that the data are sensitive to pressures ranging from a few bar up to $\sim 10^{-3}$ bar. 

Figure~\ref{fig:PT_profile} shows that the lower atmosphere is very consistent with the Sonora thermal profile of $\mathrm{T_{eff} = 2600~K}$ and $\mathrm{logg = 5.0}$, while the upper atmosphere is colder and less isothermal. In the EddySed model, the cloud base is set at the intersection of the $\mathrm{P-T}$ profile and a given cloud condensation curve. Hence, the lack of features in the emission contribution function at this intersection shows that our high-resolution data set is largely insensitive to clouds. This is expected due to our relatively small wavelength range, 2.29-2.49 $\mu$m, and due to the high temperature of \Ncomp. We also note that the clear model gives almost identical $\mathrm{P-T}$ shapes.

\section{Discussion and summary}\label{conclusion}
This paper presented the observation and characterization of the hot brown dwarf \Ncomp\ using high-resolution spectroscopic KPIC data. Using the nested-sampling technique and \pRT\, we tested different forward-retrieval models using both clear and cloudy models, and we measured some of the properties of this companion. Our results are listed below. 

\begin{itemize}
 \item By comparing the measured Bayes factors, we found that the different models we tested were consistent with each other. In addition, most of the measured parameters seem independent of assumptions about the presence or absence of clouds, nor do they rely on the accurate inference of the temperature structure. While we used the EddySed ($\mathrm{MgSiO_{3} + Fe}$, cd) model as our main model because it has the highest Bayes factor (with $\mathrm{ln(B) = 3.3}$), we found no evidence of a cloudy atmosphere for \Ncomp, which can be explained by the fact that the clouds are probably at too high a pressure in the atmosphere of \Ncomp\ to have a major impact in our high-dispersion data. For instance, no cloudy features could be observed in the emission contribution presented in Figure~\ref{fig:PT_profile}, where the cloud condensation curves of $\mathrm{MgSiO_3}$ and Fe should have intersected with our retrieved $\mathrm{P-T}$ profile. 
 
 \item From our best-fit atmospheric model, we measure an RV of $-25.0^{+0.02}_{-0.03}$~km~s$^{-1}$ and a $v\sin i$ of $12.72^{+0.03}_{-0.02}$~km~s$^{-1}$ for \Ncomp. We also found consistent metallicity values, with $[$Fe/H$] = -0.62^{+0.02}_{-0.02}$, inferring that \Ncomp\ has a substellar metallicity. This substellar metallicity seems surprising and may imply that our results are biased because the K band alone is potentially not the best band to determine the metallicity (see \citealt{Nowak2020}). This hypothesis, however, contradicts other results found in previous papers (see, e.g., \citealt{Wang2023, Xuan2024}), where the metallicity was measured to $0.2-0.3$ dex on other benchmark brown dwarfs. In future work, measurements of its metallicity at other wavelengths could be used to confirm the surprisingly low metallicity or find the source for a metallicity bias in the KPIC measurements.

 \item The C/O ratio values measured for the clear and cloudy models match the value of its host star ($0.40 \pm 0.20$). This chemical homogeneity between a companion and its host star is expected for models in which the brown dwarf companion formed via gravitational fragmentation in molecular clouds or massive protostellar disks \citep{Padoan2004, Stamatellos2007}. In addition, the high orbital eccentricity of \Ncomp\ ($\sim 0.76$) could contribute to the formation scenarios presented by some simulations, which suggested that brown dwarfs typically form in unstable multiple systems that undergo chaotic interactions \citep{Thies2010, Bate2012}. From our best-fit model, we measured a C/O ratio of $0.50\pm0.01$. This 0.01 error is the statistical precision and might not be realistic. By comparing the results obtained from different epochs, we could assess some level of systematic error (e.g., as done in \citealt{Xuan2024}). Due to the large error bar found for the C/O ratio of the host star, \Ncomp\ might still be enriched in C/O compared to \Nstar. While unlikely, this could be explained if \Ncomp\ formed close to the CO ice line from material enriched in carbon \citep{AliDib2014,Schneider2021}. 
 
 \item Finally, we detected for all models the isotopolog $^{13}$CO in the atmosphere of \Ncomp. We first found a $\mathrm{^{12}CO/^{13}CO}$ ratio of $115^{+26}_{-17}$ for the clear model, which is slightly above the carbon isotope ratio measured for the Sun ($93.5 \pm 3$, \citealt{Lyons2018}). However, the $\mathrm{^{12}CO/^{13}CO}$ ratio value measured for our best-fit cloudy model is $98^{+26}_{-19}$ which is within the 1$\sigma$ error bar of the solar value and higher than the ratio of the local interstellar medium ($\sim 68$, \citealt{Milam2005}). These results are expected for young objects (e.g., \citealt{Zhang2021a}), and by comparing our measured carbon isotopolog ratio to other objects, we might constrain possible formation mechanisms. For instance, \cite{Zhang2021b} measured a $\mathrm{^{12}CO/^{13}CO}$ ratio of $\sim 31$ for a young and wide-orbit super-Jupiter. Super-Jupiters might form via the core-accretion scenario \citep{Pollack1996, Lambrechts2012}, which can lead to $^{13}$CO enrichment through ice accretion. This would lower the $\mathrm{^{12}CO/^{13}CO}$ ratio. Hence, the higher $\mathrm{^{12}CO/^{13}CO}$ ratio value measured for \Ncomp\ could favor another formation scenario in which the brown dwarf is formed via gravitational collapse or disk instability. This hypothesis is corroborated by our observations of the C/O ratio. 
 \end{itemize}

From these results, we can infer the likely formation mechanism of \Ncomp, namely that this hot brown dwarf probably formed via turbulent fragmentation in a molecular cloud or disk instability. This conclusion would confirm the trend of chemical homogeneity observed between brown dwarf companions (with masses $\sim13-70\mathrm{M_{Jup}}$) and their host stars (e.g., \citealt{wang2022}, \citealt{Xuan2022}, \citealt{Xuan2024b}, \citealt{Hsu2024}). 

However, our analysis revealed that not all of the parameters were properly constrained and pointed out some of the limits of this study. By only using a small wavelength range (between 2.29 and 2.49~$\mu$m), we did  not fully constrain some bulk properties of \Ncomp\ , such as some of the cloudy parameters and the radius of the companion. To better constrain the cloud properties and abundances, it would be important to consider a joint analysis with available low-resolution spectroscopy data or with high-resolution spectroscopy in other spectral bands, such as the H band with the Rigorous Exoplanetary Atmosphere Characterization with High dispersion coronography (REACH; \citealt{Kotani2020}) or the High-Resolution Imaging and Spectroscopy of Exoplanets (HiRISE; \citealt{Vigan2024}). The addition of mid-infrared spectroscopy from the James Webb Space Telescope will also bring major insight in the coming years for our understanding of the atmosphere of substellar companions \citep[e.g.,][]{Carter2023,Miles2023}.

\begin{acknowledgements}
JC and AV acknowledge funding from the European Research Council (ERC) under the European Union's Horizon 2020 research and innovation programme (grant agreement No. 757561). Funding for KPIC has been provided by the California Institute of Technology, the Jet Propulsion Laboratory, the Heising-Simons Foundation (grants \#2015-129, \#2017-318, \#2019-1312, \#2023-4598), the Simons Foundation, and the NSF under grant AST-1611623.
\end{acknowledgements}

\bibliographystyle{aa}
\bibliography{biblio}

\onecolumn
\begin{appendix}
\section{Posteriors from our retrieval}
\begin{figure}[!b]
    \includegraphics[width=1\textwidth]{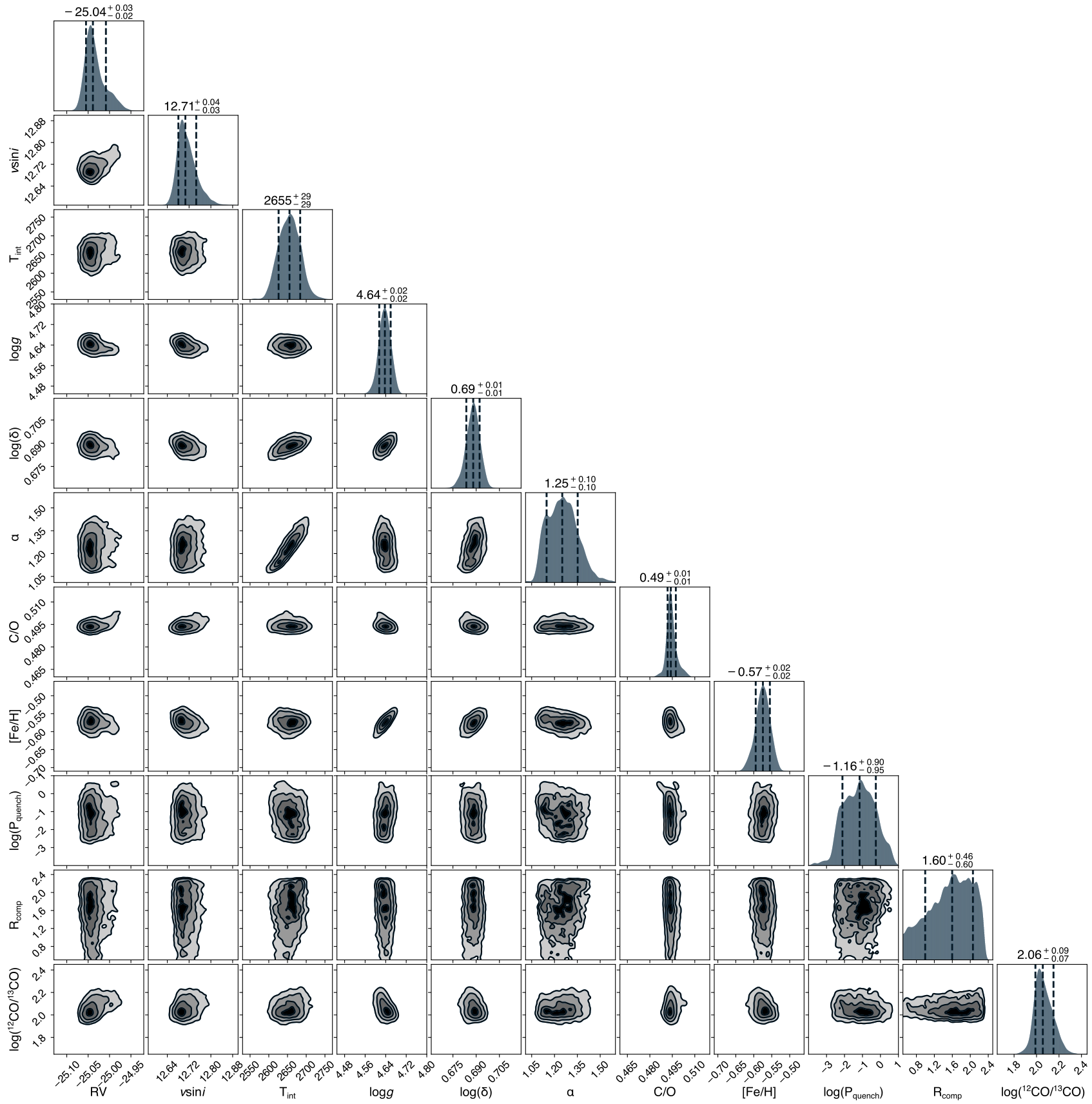}
    \caption{\label{fig:Corner_2} Extended corner plot showing the posterior distribution for our clear model retrieval for \Ncomp\ (see Section~\ref{clear model}). The titles at the top of each histogram show the median and 1$\sigma$ error.}
\end{figure}


\begin{figure*}
    \centering
    \includegraphics[width=1\textwidth]{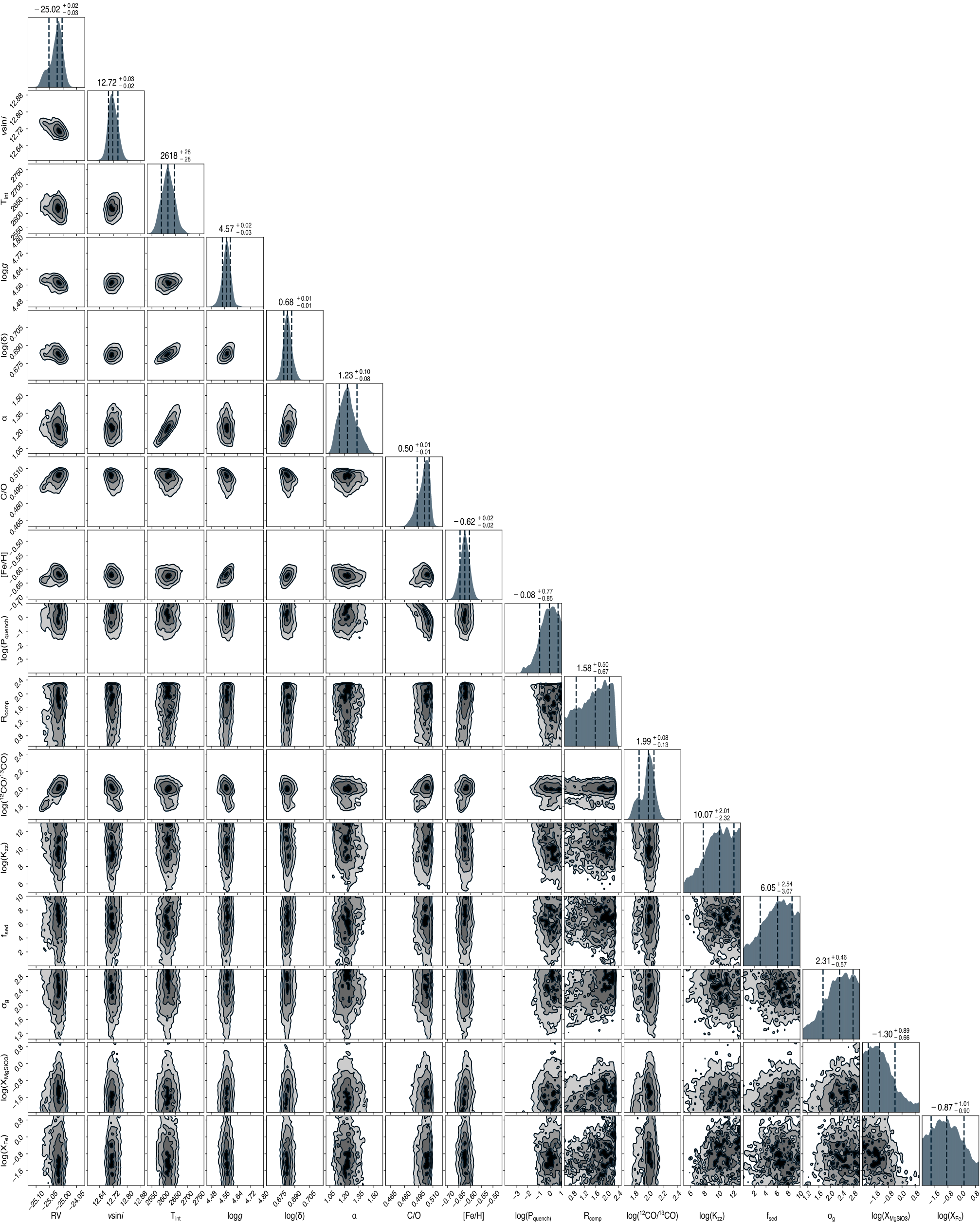}
    \caption{\label{fig:Corner_3} Extended corner plot showing the posterior distribution for our best-fit cloudy model retrieval EddySed (MgSiO3 + Fe, cd) for \Ncomp\ (see Section~\ref{cloudy model}). The titles at the top of each histogram show the median and the 1$\sigma$ error.}
\end{figure*}
\end{appendix}%

\end{document}